\documentclass[aps,prd,twocolumn,nofootinbib,showpacs,superscriptaddress]{revtex4-1}

\usepackage{amsfonts}
\usepackage{amsmath}
\usepackage{amssymb}
\usepackage{bm}
\usepackage{dcolumn}
\usepackage{epsfig}
\usepackage{graphicx}
\usepackage{graphics}
\usepackage[latin1]{inputenc}
\usepackage{latexsym}
\usepackage{rotating}
\usepackage{hyperref}
\usepackage[x11names]{xcolor}
\usepackage{xspace} 
\usepackage{mathrsfs}
\usepackage{subfigure}
\usepackage{enumerate}
\usepackage{tabularx}
\usepackage{multirow,hhline}
\usepackage{booktabs}
\usepackage{siunitx}
\usepackage{array}
\newcolumntype{L}[1]{>{\raggedright\let\newline\\\arraybackslash\hspace{0pt}}m{#1}}
\newcolumntype{C}[1]{>{\centering\let\newline\\\arraybackslash\hspace{0pt}}m{#1}}
\newcolumntype{R}[1]{>{\raggedleft\let\newline\\\arraybackslash\hspace{0pt}}m{#1}}
\newlength{\Oldarrayrulewidth}

\hypersetup{
	colorlinks=true,
	linkcolor=magenta,
	filecolor=blue,
	citecolor = cyan,
	urlcolor=cyan,
}
\usepackage{ulem}
\definecolor{rufous}{rgb}{0.66, 0.11, 0.03}
\graphicspath{{../Figures/}}

\newcommand{\be}{\begin{equation}}
\newcommand{\ee}{\end{equation}}
\newcommand\ba{\begin{eqnarray}}
\newcommand\bse{\begin{subequations}}
\newcommand\ea{\end{eqnarray}}
\newcommand\ese{\end{subequations}}
\newcommand{\bwt}{\begin{widetext}}
\newcommand{\ewt}{\end{widetext}}

\newcommand{\eq}{\,=\,}
\newcommand{\mg}[1]{\left|#1\right|}

\newcommand{\mat}{{\mbox{\tiny mat}}}

\newcommand{\DEF}{{\mbox{\tiny DEF}}}
\newcommand{\MO}{{\mbox{\tiny MO}}}

\newcommand{\pbdot}{\dot{P}_b}
\newcommand{\omegadot}{\dot{\omega}}

\newcommand{\ppN}{{\mbox{\tiny PPN}}}

\newcommand{\intr}{{\mbox{\tiny int}}}
\newcommand{\obs}{{\mbox{\tiny obs}}}

\newcommand{\mon}{{\mbox{\tiny mon}}}
\newcommand{\dip}{{\mbox{\tiny dip}}}
\newcommand{\quadr}{{\mbox{\tiny quad}}}
\newcommand{\effective}{1 + \alpha_A \alpha_B}
\newcommand{\kep}{\left(\dfrac{G_{AB} M n}{c^3}\right)}
\newcommand{\mtext}[1]{{\mbox{\tiny #1}}}
\newcommand{\msun}{M_\odot}
\newcommand{\theory}{{\mbox{\tiny th}}}

\definecolor{red(ncs)}{rgb}{0.77, 0.01, 0.2}

\begin{document}

\title{Binary Pulsar constraints on massless scalar-tensor theories using Bayesian statistics}

\author{David Anderson}
\affiliation{eXtreme Gravity Institute, Department of Physics, Montana State University, Bozeman, MT 59717, USA.}

\author{Paulo Freire}
\affiliation{Max-Planck-Institut f\"ur Radioastronomie, Auf dem H\"{u}gel 69, D-53121 Bonn, Germany}

\author{Nicol\'as Yunes}
\affiliation{eXtreme Gravity Institute, Department of Physics, Montana State University, Bozeman, MT 59717, USA.}

\date{\today}

\begin{abstract}\label{sec:abstract}

Binary pulsars provide some of the tightest current constraints on modified theories of gravity and these constraints will only get tighter as radio astronomers continue timing these systems. These binary pulsars are particularly good at constraining scalar-tensor theories in which gravity is mediated by a scalar field in addition to the metric tensor. Scalar-tensor theories can predict large deviations from General Relativity due to the fact that they allow for violation of the strong-equivalence principle through a phenomenon known as scalarization. This effect appears directly in the timing model for binary pulsars, and as such, it can be tightly constrained through precise timing. In this paper, we investigate these constraints for two scalar-tensor theories and a large set of realistic equations of state. We calculate the constraints that can be placed by saturating the current $1\sigma$ bounds on single post-Keplerian parameters, as well as employing Bayesian methods through Markov-Chain-Monte-Carlo simulations to explore the constraints that can be achieved when one considers all measured parameters simultaneously. Our results demonstrate that both methods are able to place similar constraints and that they are both indeed dominated by the measurements of the orbital period decay. The Bayesian approach, however, allows one to simultaneously explore the posterior distributions of not only the theory parameters but of the masses as well.

\end{abstract}

\maketitle
\section{Introduction}\label{ch5:sec:introduction}

General Relativity (GR) has been the most successful theory of gravity over the last century, passing all current tests with great success. Our ability to test gravitational theories is becoming even stronger~\cite{Yunes:2013dva,Yunes:2016jcc} with the recent observations of gravitational waves (GW)~\cite{Abbott:2016blz, Abbott:2017oio,TheLIGOScientific:2017qsa} and the continued monitoring of binary pulsars (PSRs)~\cite{Freire:2012mg, Kramer:2009zza, Kramer:2016kwa}. As more neutron star (NS) systems are discovered through gravitational wave observations we will be able to pin down the nuclear equation of state (EOS) and other NS properties~\cite{Abbott:2018exr, Abbott:2018wiz}. Such observations can help break the degeneracy that arise between EOS effects and those introduced by modified theories of gravity.

Even with the success of GR, there are many theoretical reasons to consider modified theories. Some of the most well-motivated modified theories are scalar-tensor theories (STTs) of gravity, in which gravity is mediated by the metric tensor \emph{and} a dynamical scalar field. These theories are defined by a choice of conformal coupling function that dictates the manner in which the scalar field couples to matter, and ultimately the level at which the strong-equivalence principle (SEP) is violated. Originally proposed by Jordan~\cite{Jordan-book, Jordan:1959eg}, Fierz~\cite{Fierz:1956zz}, Brans~\cite{Brans:1961sx}, and Dicke (JFBD) as some of the most natural alternatives to GR, STTs were later extended by Damour and Esposito-Far\'ese (DEF) to include higher order terms in the conformal coupling, as well as multiple scalar fields~\cite{Damour:1992we}. Recently, a slight variation on this theory, proposed by Mendes and Ortiz (MO)~\cite{Mendes:2016fby}, has gained some attention as it arises from more fundamental considerations.

While Solar System observations have the ability to tightly constrain STTs, these weak field constraints do not always translate to tight restrictions in the strong field regime. In particular, STTs give rise to a phenomenon in NSs known as scalarization in which the scalar field can become excited far above its background (weak field) value. Neutron stars in STTs can acquire a so-called scalar charge which quantifies the $1/r$ behavior of the scalar field in a far field expansion from the NS. Such modifications to NS spacetimes directly affect observables, particularly those that can be probed with binary PSR observations. The scalar charges that NSs can develop appear directly in the timing model that is used to predict when we should observe pulses from binary PSRs. Astronomers are able to time PSRs so precisely that even slight deviations from the GR predictions are highly constrained, thus providing some of the best constraints on STTs.

Typically, constraints on STTs are placed through the observed rate of decay of the orbital period of the binary. STTs predict dipolar gravitational radiation, which enters at lower post Newtonian (PN) order relative to the typical quadrupole radiation predicted by GR, and thus, speeds up the orbital decay rate. Since observations seem to suggest the absence of this extra dipole effect, STTs can be constrained from the observation of the orbital decay rate with precision that can exceed that of Solar System observations. However, STTs predict that \emph{all} post-Keplerian (PK) parameters~\cite{Damour:1991rd} are modified from the GR prediction. Thus, observations of post-Keplerian parameters other than the orbital period decay can in principle be combined to allow for even tighter constraints.

In this paper we investigate the type of constraints that can be placed on STTs, particularly the theories proposed by DEF and MO, by using multiple post-Keplerian parameters, as well as multiple PSR systems. We use a combination of PSR-white-dwarf systems and PSR-NS systems with a wide range of PSR masses to explore the effects of scalarization and place constraints in a self-consistent manner. We also explore the effects of the NS EOS and how it affects the strength of such constraints. Such studies are carried out with Bayesian methods through Markov-Chain Monte-Carlo simulations that explore the \emph{full} parameter space and determine which values of the STT parameters are most consistent with observations. This approach allows us to accurately take into account any correlations between observed post-Keplerian parameters and ensures that the constraints we place are self-consistent.


The study and results summarized above are a comprehensive extension of other work that also employed a Markov-Chain Monte-Carlo methodology for binary pulsar tests of GR. In 2017,~\cite{Shao:2017gwu} carried out a Markov-Chain Monte-Carlo exploration of the posterior probability distribution of the two main STT parameters that characterize DEF theory. This analysis used 5 PSRs that currently have good measurements of the component masses (either through optical counterparts or through measurements of the PPK parameters associated with the Shapiro time delay) and of one additional PPK parameter (the orbital period decay). A Bayesian analysis was then employed to obtain posteriors for the STT parameters for 11 different EOSs, each marginalized over the masses. 

Our work differs from this analysis in various ways. 
First, we do not marginalize only over the component masses, but also over the equations of state (treating the latter as a discrete parameter). 
Second, we analyze binaries with measurements of any PPK parameter and not just the orbital period decay, which is possible due to a prior deep exploration of scalar charges~\cite{david-private}. 
Third, we greatly extend the exploration of the STT parameter space, not only considering the region where spontaneous scalarization activates. 
Fourth, we do not only focus on the DEF theory, which is already stringently constrained by cosmological and Solar System observations~\cite{Sampson:2014qqa,Anderson:2016aoi,Anderson:2017phb}, but we also study the MO theory.   

Using the data for scalar charges provided in~\cite{david-private}, we place constraints on DEF and MO theory using all measured post-Keplerian parameters available for a set of 7 PSRs with 11 different EOSs. We show that the EOS has relatively little impact on these types of parameters and does not change the relative strength of any constraints placed from different PSRs. We find that other post-Keplerian parameters beyond the orbital period decay, like the rate of periastron advance, are, in some cases, able to place tighter constraints on STTs for PSR-NS systems. Finally, when we take a Bayesian approach to place constraints, we find similar results to what one finds when using the approach taken in the past~\cite{Esposito-Farese:2011cha,Freire:2012mg}, except that a Bayesian approach allows us to (i) easily stack all constraints from all PSR observations and (ii) easily marginalize over masses and EOSs.  

We start with a description of STTs and the timing model in Sec.~\ref{ch5:sec:background} in order to lay down the foundation for the rest of the paper. Here we discuss the current constraints on STTs from Solar System observations, as well as the details of the various post-Keplerian parameters appearing in the timing model. In Sec.~\ref{ch5:sec:pulsars} we investigate the various constrains that can be placed on STTs, first through the standard method that is employed in the literature, and then through MCMC methods in \S\ref{ch5:sec:pulsars:bayesian}. We then conclude in Sec.~\ref{ch5:sec:conclusion} with a discussion of our results and future work. Throughout this paper, we follow the conventions of~\cite{Misner:1974qy}. 


\section{Timing binary PSRs in STTs}\label{ch5:sec:background}

In this section we will introduce the basics of STTs and the post-Keplerian parameters that appear in the timing model. We will discuss the particular theories we consider in this paper and the current constraints that Solar System observations place on these theories. We give a brief discussion of the scalar charges and how they are calculated, and discuss their importance for timing binary PSRs. For completeness we provide a summary of the timing model and present the various post-Keplerian parameters in the context of STTs.


\subsection{Scalar-tensor theories}\label{ch5:sec:backgroun:field_equations}

In this paper we will focus explicitly on massless STTs in which there exists a single scalar field non-minimally coupled to the metric tensor $g_{\mu\nu}$. These theories can be derived from an action in the so called \emph{Einstein frame} given by~\cite{Damour:1992we, Damour:1993hw, Damour:1996ke}
\be
S \eq \int \dfrac{d^4x}{c} \dfrac{\sqrt{-g}}{4\kappa}\left[ R - 2g^{\mu\nu} \partial_\mu \varphi \partial_\nu \varphi\right] + S_\mat\left[\chi, A^2(\varphi)g_{\mu\nu}\right]\,\,,
\label{ch5:eq:einstein-action}
\ee
where $g$ and $R$ are the determinant and Ricci scalar associated with the metric $g_{\mu\nu}$, $\kappa = 4 \pi G/c^4$, $\chi$ are any matter fields, and $A(\varphi)$ is a conformal factor that determines how the scalar field $\varphi$ couples to matter. 

The field equations resulting from variation of action above with respect to $g_{\mu\nu}$ and $\varphi$ are given by
\ba
R_{\mu\nu} &\eq& 2 \partial_\mu \varphi \partial_\nu \varphi + 2\eta \left(T_{\mu\nu}^\mat - \dfrac{1}{2}g_{\mu\nu} T^\mat\right)\,\,,
\label{ch5:eq:einstein-equation}\\
\Box \varphi &\eq& -\kappa \alpha(\varphi)T^\mat\,\,,
\label{ch5:eq:kg-equation}
\ea
where the stress-energy tensor is defined by
\be
T_{\mu\nu}^\mat \equiv \dfrac{2c}{\sqrt{-g}}\left(\dfrac{\delta S_m}{\delta g^{\mu\nu}}\right)\,\,,
\label{ch5:eq:SET-definition}
\ee
and $T^\mat \equiv g_{\mu\nu} T_{\mu\nu}^\mat$ is its trace. 

In addition to the $A(\varphi)$ function that appears in the action, it is also convenient to introduce a couple of additional quantities. Let us then define
\be
\alpha(\varphi) \eq \dfrac{\partial \ln A(\varphi)}{\partial \varphi}\,\,,
\label{ch5:eq:conformal_coupling}
\ee
which we designate the conformal coupling, as it appears directly in the equations of motion for the scalar field. Another quantity that appears directly in the parameterized-post-Newtonian (PPN) and parameterized-post-Keplerian (PPK) formalisms is the derivative of this quantity, namely,
\be
\beta(\varphi) \eq \dfrac{\partial \alpha(\varphi)}{\partial \varphi}\,\,,
\label{ch5:eq:conformal_curvatuve}
\ee
and quantifies the non-linear behavior of $A(\varphi)$.

A particular STTs is defined by one's choice of the conformal factor $A(\varphi)$, or likewise $\alpha(\varphi)$, and will ultimately determine how the scalar field reacts to the presence of matter, or likewise the gravitational potential generated by such matter. In this paper, we consider two models that exhibit similar behavior when $\varphi \ll 1$ but allows for much different behavior elsewhere. The first model we consider is DEF theory~\cite{Damour:1993hw, Damour:1992we}, which is defined by
\ba
A(\varphi) &\eq& e^{\beta_0 \varphi^2 /2}\,\,,
\label{ch5:eq:DEF-conformal}\\
\alpha(\varphi) &\eq& \beta_0 \varphi\,\,,
\label{ch5:eq:DEF-coupling}\\
\beta(\varphi) &\eq&  \beta_0\,\,,
\label{ch5:eq:DEF-curvature}
\ea
where $\beta_0$ enters directly as a free parameter. The other model we consider is MO theory~\cite{Mendes:2016fby}, which is defined by
\ba
A(\varphi) &\eq& \left[\cosh\left(\sqrt{3}\, \beta_0 \varphi\right)\right]^{1/(3\beta_0)}\,\,,
\label{ch5:eq:COSH-conformal}\\
\alpha(\varphi) &\eq& \dfrac{\tanh\left(\sqrt{3}\,\beta_0 \varphi\right)}{\sqrt{3}}\,\,,
\label{ch5:eq:COSH-coupling}\\
\beta(\varphi) &\eq&  \beta_0 \,\text{sech}^2 \left( \sqrt{3}\,\beta_0 \varphi \right)\,\,,
\label{ch5:eq:COSH-curvature}
\ea
where, again, $\beta_0$ enters directly as a free parameter. 

Both theories are subject to a boundary condition at spatial infinity such that the scalar field has a cosmologically determined background value $\varphi_\infty$. This quantity then becomes another free parameter of these theories and must be chosen to satisfy weak-field constraints, like those from Solar System observations. However, we choose the parameterization $\varphi_\infty = \alpha_0/\beta_0$ and let $\alpha_0$ become our second free parameter appearing in both theories. A more detailed description of this parameterization of STTs can be found in Refs.~\cite{david-private, Damour:2007uf}. It is worth pointing out, however, the this parameterization enforces certain relations when one inserts $\varphi_\infty$ into Eqs.~\eqref{ch5:eq:DEF-coupling}-\eqref{ch5:eq:DEF-curvature}, namely
\ba
\alpha_\infty^{\DEF} &\eq& \alpha(\varphi_\infty) \eq \alpha_0\,\,,
\label{ch5:eq:alpha_bar-DEF}\\
\beta_\infty^{\DEF} &\eq& \beta(\varphi_\infty) \eq\beta_0\,\,,
\label{ch5:eq:beta_bar-DEF}
\ea
in DEF theory, and when using Eqs.~\eqref{ch5:eq:COSH-coupling}-\eqref{ch5:eq:COSH-curvature}
\ba
\alpha_\infty^{\MO} &\eq& \tanh\left( \sqrt{3}\,\alpha_0\right)/\sqrt{3}\,\,,
\label{ch5:eq:alpha_bar-MO}\\
\beta_\infty^{\MO} &\eq& \beta_0\,\text{sech}^2 \left( \sqrt{3}\,\alpha_0\right)\,\,,
\label{ch5:eq:beta_bar-MO}
\ea
in MO theory.

The quantities in Eqs.~\eqref{ch5:eq:alpha_bar-DEF}-\eqref{ch5:eq:beta_bar-MO} appear directly in the weak field predictions made by these theories. The choice of parameters $(\alpha_0,\,\beta_0)$ will determine the local value of the gravitational constant via the relation
\be
G_N \eq G\left[A^2_\infty\left(1+ \alpha^2_\infty\right)\right]\,\,,
\label{ch5:eq:G-newton}
\ee
as well as the PPN parameters $\gamma_\ppN$ and $\beta_\ppN$~\cite{Will:1993ns, Will:2014kxa}. The former is given by
\be
\bar{\gamma} \eq \mg{1 - \gamma_\ppN} \eq \dfrac{2 \alpha_\infty^2}{1 + \alpha_\infty^2}\,\,,
\label{ch5:eq:ppn-gamma}
\ee
while the latter is given by
\be
\bar{\beta} \eq \mg{1 - \beta_\ppN} \eq \dfrac{|\beta_\infty|\alpha_\infty^2}{2\left(1 + \alpha_\infty^2\right)^2}\,\,,
\label{ch5:eq:ppn-beta}
\ee
The $\gamma_{\ppN}$ parameter is a measure of the spatial curvature induced by a unit rest mass and has been measured from the Shapiro time delay observed by the Cassini spacecraft~\cite{Bertotti:2003rm, Will:2014kxa}, being constrained to $\mg{1 - \gamma_\ppN} \lesssim 2.3 \times 10^{-5}$. The $\beta_{\ppN}$ parameter is a measure of the amount of non-linearity in the superposition law for gravity, and it is constrained from observations of the perihelion shift of Mercury~\cite{Will:2014kxa} to be $\mg{1 - \beta_\ppN} \lesssim 8\times 10^{-5}$. From these relations we infer constraints on $\alpha_\infty$ and $\beta_\infty$, namely
\be
\alpha_\infty^2 \lesssim \dfrac{\bar{\gamma}}{2 - \bar{\gamma}}\,\,,
\label{ch5:eq:alpha_inf1}
\ee
from Eq.~\eqref{ch5:eq:ppn-gamma}, and
\be
\alpha_\infty^2(\beta_0) \lesssim \dfrac{|\beta_\infty|}{\bar{\beta}} - 1 - \dfrac{1}{4 \bar{\beta}} \sqrt{|\beta_\infty| (|\beta_\infty| - 8 \bar{\beta}})\,\,,
\label{ch5:eq:alpha_inf2}
\ee
from Eq.~\eqref{ch5:eq:ppn-beta}, which only places tighter constraints for $|\beta_\infty| \gtrsim 14$. 

\subsection{Scalarization of neutron stars}\label{ch5:sec:background:scalarization}

While STTs can be constrained rather tightly from Solar System observations they are still able to produce significant deviations from GR near strongly self-gravitating matter like NSs. These strong field effects are a result of a well-studied phenomenon known as scalarization~\cite{Damour:1993hw, Damour:1992we, Damour:1996ke} . When the coupling parameter $\beta_0$ is sufficiently negative ($\lesssim -4.3$) the scalar field can grow rapidly inside the NS even when the cosmologically determined background value of the field approaches zero. This is precisely how STTs satisfy Solar System constraints but produce observable deviations from GR in the strong field regime.

The effects of scalarization can be quantified by quantities known as scalar charges and they enter directly into the timing model presented in the next section. There are three scalar charges that appear, the first of which is defined via
\be
\alpha_A \eq \left.\dfrac{\partial \ln m_A}{\partial \varphi_\infty}\right|_{\bar{m}_A}\,\,,
\label{ch5:eq:charge1}
\ee
for the $A$th NS of a system and where the derivative must be taken with the baryonic mass held constant. This scalar charge measures the ``sensitivity'' of the NS's mass to variations in the background scalar field, and thus, it represents an effective coupling between the NS and the scalar field. This quantity is the strong field counterpart to the weak field parameter $\alpha_\infty$. The second relevant scalar charge is the strong field equivalent to $\beta_\infty$, and thus, it is a derivative of Eq.~\eqref{ch5:eq:charge1}, i.e.
\be
\beta_A \eq \left.\dfrac{\partial \alpha_A}{\partial \varphi_\infty}\right|_{\bar{m}_A}\,\,,
\label{ch5:eq:charge2}
\ee
where again the baryonic mass must be held constant. This scalar charge encodes non-linear interactions between the binary component. The final scalar charge of interest is linked to the NS's moment of inertia $I_{A}$, and it is defined as
\be
k_A \eq \left.\dfrac{\partial \ln I_A}{\partial \varphi_\infty}\right|_{\bar{m}_A}\,\,,
\label{ch5:eq:charge3}
\ee
with the baryonic mass once more held fixed. Similar to the relations between $\alpha_A$ and the NS's mass, $k_A$ quantifies the sensitivity of the NS's moment of inertia to the scalar field. This quantity becomes most relevant when a NS is in a binary system with another NS. As the NSs orbit one another, they will move through each other's gravitational potential, effectively altering the local value of the scalar field. The NS's moment of inertia will then fluctuate throughout the orbit and produce an additional time delay that is measurable in principle.

Calculating these scalar charges is not very difficult but can be numerically expensive over large regions of the $(\alpha_0, \,\beta_0)$ parameter space. For the Bayesian methods used in this paper, it would be nearly impossible to compute these quantities as often as they would be needed for millions of likelihood calculations. However, these quantities have recently been calculated over a large region of parameter space and for a wide range of EOS~\cite{david-private}, consistent with the PSR mass in J0348+0432 and the recent constraints placed from LIGO and VIRGO~\cite{Abbott:2018exr}. In this paper, we make use of this data in order to avoid solving for the scalar charges on the fly during our MCMC simulations. 


\subsection{Timing model in STTs}\label{ch5:sec:background:timing}

In order to use binary PSRs to place constraints on a gravitational theory, one must work out how that theory predicts modifications to the motion of a binary system. Therefore, one must develop a timing formula that captures the relativistic effects of the theory by relating the observed time of arrival (TOA) of the pulse to the time the pulse was emitted. Blandford and Teukolsky (BT)~\cite{Blandford1976ApJ} originally developed such a formula to explain the, then recent, discovery by Hulse and Taylor of the the first binary PSR B1913+16~\cite{Hulse:1974eb}. Later, Damour and Deruelle (DD) developed a full 1PN description of the two-body problem in a way that allowed for a parameterized description of the timing model, encapsulating all relativistic effects in GR. This model was then extended by Damour and Taylor~\cite{Damour:1991rd} in a phenomenological manner such that the DD model could be used generically to constrain any conservative theory of gravity. The parameters of the Damour-Taylor model are the two masses ($m_A$ and $m_B$) of the binary, the standard Keplerian orbital parameters $\{ P_b, n=2\pi/P_b,\,T_0,\,\omega_0,\, e_0,\,x_0 \}$ (orbital period, orbital frequency,  time of periastron passage, location of periastron, eccentricity, and projected semi-major axis), and a set of post-Keplerian parameters. In this section we will review the timing formula of DD and present the relevant post-Keplerian parameters of Damour and Taylor in the context of scalar-tensor theories.

The timing model is traditionally (BT and DD) written as
\be
D \tau_a \eq T_e + \Delta_R + \Delta_E + \Delta_S + \mathcal{O}(c^{-4})\,\,,
\label{ch5:eq:timing_model}
\ee
where $\tau_a$ is the infinite frequency barycenter arrival time, $T_e$ is the proper time of emission, and $D$ is a Doppler factor accounting for center of mass motion between the binary and the Solar System barycenter. Using $\tau_a$ here means that one must have accurately accounted for the time delays associated with dispersion from the interstellar medium and corrections that transform this barycenter arrival time to the time kept at the observatory on Earth's surface~\cite{Edwards:2006zg, lorimer2005handbook}. The Doppler factor $D$ is an inconsequential constant that can be transformed out of the timing formula through a redefinition of units, and restored later if needed~\cite{Damour:1991rd}. The last three terms in Eq.~\eqref{ch5:eq:timing_model} account for all other time delays occurring between the binary and the Solar System barycenter. The R\"{o}mer delay $\Delta_R$ is simply the classical light travel across the binary and depends on the PSR's position in its orbit. The magnitude of this time delay is determined by the projected semi-major axis along the line of sight.

The Einstein delay $\Delta_E$ arises from relating coordinate time of emission to the proper time of emission, and thus, it depends on the metric component $g_{00}$. This time delay takes the form
\be
\Delta_E \eq \gamma \sin u \,\,,
\label{ch5:eq:einstein_delay}
\ee
where $\gamma$ is a post-Keplerian parameter and $u$ is an eccentric anomaly like variable that is the function of $T_e$ found by solving 
\be
u - e \sin u \eq n\left[(T_e - T_0) - \dfrac{\pbdot}{2p_b}(T_e - T_0)^2\right]\,\,.
\label{ch5:eq:kpeler_equation}
\ee
with $\pbdot$ being another post-Keplerian parameter describing the change in orbital period, which we will discuss later. The coefficient $\gamma$ is determined when one accounts for the gravitational redshift of the companion and the second-order Doppler shift from the PSR's motion in the line element, using Kepler's equations to integrate. 

There is also an additional effect described by the Einstein delay that is related to violation of the SEP.  Theories that violate the SEP will allow the PSR's moment of inertia to change throughout the orbit as it moves through the gravitational potential of the companion. This means that the PSR's rotational frequency will vary throughout its orbit in order to conserve angular momentum. Such effects add another time delay to the timing formula and are directly proportional to $\sin u$, and therefore, they are included in the definition of $\gamma$~\cite{Will:1993ns, Damour:1991rd}. 

The final time delay appearing in Eq.~\eqref{ch5:eq:timing_model} is the Shapiro delay $\Delta_S$ resulting from the fact that light must travel in the curved background of the binary. This term can be thought of as the first relativistic correction to $\Delta_R$ and is most easily measured when one observes the binary edge on. The magnitude of the Shapiro delay is denoted by a post-Keplerian parameter $r$ and is labeled the ``range'' of the Shapiro delay. Another post-Keplerian parameter, $s \equiv \sin \iota$, representing the ``shape'' of the Shapiro delay, and enters here to characterize how this time delay is effected by the orbit's inclination relative to the line of sight. 

While Eq.~\eqref{ch5:eq:timing_model} captures the delays associated with the pulse traveling in the curved spacetime between the PSR and the Solar System barycenter, it does not account for any secular variations in the Keplerian parameters of the binary. Thus, in principle, one must also account for three other post-Keplerian parameters, i.e. $\dot{x}$, $\dot{e}$, and $\dot{\omega}$, via 
\ba
x \eq x_0 + \dot{x}(T_e - T_0)\,\,,\\
e \eq e_0 + \dot{e}(T_e - T_0)\,\,,\\
\omega \eq \omega_0 + \omegadot(T_e - T_0)\,\,,
\ea
and insert these back into the timing formula of Eq.~\eqref{ch5:eq:timing_model}. In practice, however, the rate of periastron advance $\omegadot$ is usually the only one of these parameters that can be measured consistently and its theoretical value can be determined by the methods in~\cite{DD1, Damour:1991rd}.

Thus far, we have worked in a theory independent framework and we have parameterized the various time delays with a set of measurable post-Keplerian parameters $\{\omegadot,\,\gamma,\,r,\,s \}$. In GR, these four parameters take on simple forms that are a function of the two masses and the Keplerian parameters. In other theories of gravity, however, one must determine their functional form using the formalism in~\cite{Damour:1991rd}. Thus, in the context of STTs~\cite{Damour:1991rd,Damour:1992we,Damour:2007uf, Horbatsch:2011nh}, the post-Keplerian parameters that are typically measured are given by
\begin{widetext}
\ba
\gamma &\eq& \dfrac{e}{n} \dfrac{X_B}{1+\alpha_A\alpha_B}\left(\dfrac{G_{AB} M n}{c^3}\right)^{2/3} \left[X_B(1+\alpha_A \alpha_B) + 1 + \alpha_B k_A\right]\,\,,
\label{ch5:eq:ppk_gamma}\\
\dot{\omega} &\eq& \dfrac{3n}{1 - e^2}\left(\dfrac{G_{AB} M n}{c^3}\right)^{2/3}\left[\dfrac{1 - \alpha_A \alpha_B / 3}{1 + \alpha_A \alpha_B} - \dfrac{X_A \beta_B \alpha_A^2 + X_B \beta_A \alpha_B^2}{6(1 + \alpha_A \alpha_B)^2}\right]\,\,,
\label{ch5:eq:ppk_omdot}\\
r &\eq& \dfrac{G(1 + \alpha_\infty \alpha_B) m_B}{c^3}\,\,,
\label{ch5:eq:ppk_r}\\
s &\eq& \dfrac{n a \sin \iota}{c X_B}\left(\dfrac{G_{AB} M n}{c^3}\right)^{-1/3}\,\,,
\label{ch5:eq:ppk_s}
\ea
\end{widetext}
in which we have made use of the notation in~\cite{Damour:1991rd}, where $M$ is the total mass and $X_{A,B} = m_{A,B}/M$. One will notice that the scalar charge $( \alpha_A,\,\beta_A,\,k_A)$ described earlier appear directly in these definitions.

The effects we have incorporated thus far do not capture radiative effects such as gravitational wave emission. The orbital period derivative, $\pbdot$, accounts for these radiative effects, along with others, and in general has the contributions~\cite{Damour:1991rd}
\be
\pbdot \eq \pbdot^\mtext{Acc} + \pbdot^\mtext{Shk} + \pbdot^{\dot{M}} + \pbdot^T + \pbdot^{\dot{G}} + \pbdot^\intr\,\,.
\ee
The quantity $\pbdot^\mtext{Acc}$ is due to the relative acceleration between the binary and the Solar System barycenter along the line of sight. The quantity $\pbdot^\mtext{Shk}$ is the so-called Shklovskii effect and is due to centrifugal acceleration between the two binary components. The quantity $\pbdot^{\dot{M}}$ accounts for mass loss of the system, and finally $\pbdot^T$ accounts for any tidal effects. The term $\pbdot^{\dot{G}}$ accounts for a possibly varying gravitational constant which generally happens on cosmological time scales, if for example the scalar field evolves in a cosmological potential and influences the gravitational constant. 

The intrinsic orbital period derivative, $\pbdot^\intr$, is of most interest for constraining STTs and can be found directly from the orbital energy that is lost due to gravitational radiation. In the context of STTs, $\pbdot^\intr$ has the following contributions
\be
\pbdot^\intr \eq \pbdot^{\varphi,\mon} + \pbdot^{\varphi,\dip} + \pbdot^{\varphi,\quadr}+ \pbdot^{g,\quadr}\,\,.
\label{ch5:eq:pbdot_intrinsic}
\ee
Each of these pieces can be calculated from a multipole expansion of the energy fluxes for the scalar field and the metric, which yield~\cite{Damour:1992we}
\begin{widetext}
\ba
\pbdot^{\varphi,\mon} &\eq& -\dfrac{3 \pi X_A X_B}{\effective}\kep^{5/3} \dfrac{e^2(1 + e^2 /4)}{(1-e^2)^{7/2}} \left[\dfrac{5}{3} (\alpha_A + \alpha_B) - \dfrac{2}{3}(\alpha_A X_A + \alpha_B X_B) + \dfrac{\beta_A \alpha_B + \beta_B \alpha_A}{\effective}\right]^2 \,\,,
\label{ch5:eq:pbdot_monopole}\\
\pbdot^{\varphi,\dip} &\eq& - \dfrac{2 \pi X_A X_B}{\effective} \kep \dfrac{(1 + e^2 / 2}{(1-e^2)^{7/2}}(\alpha_A - \alpha_B)^2 \,\,,
\label{ch5:eq:pbdot_dip}\\
\pbdot^{\varphi,\quadr} &\eq& - \dfrac{32 \pi X_A X_B}{5 ( \effective)} \kep^{5/3} (1 - e^2)^{-7/2} \left(1 + \dfrac{73}{24}e^2 + \dfrac{37}{96}e^4\right)\left[X_B \alpha_A + X_A \alpha_B\right] \,\,,
\label{ch5:eq:pbdot_quad_phi}\\
\pbdot^{g,\quadr} &\eq& - \dfrac{192 \pi X_A X_B}{5 ( \effective)} \kep^{5/3} (1 - e^2)^{-7/2} \left(1 + \dfrac{73}{24}e^2 + \dfrac{37}{96}e^4\right)\,\,.
\label{ch5:eq:pbdot_quad_GR}
\ea
\end{widetext}
In GR only the quadrupolar term survives, but in STTs, both monopole and dipole radiation exist as well, the latter entering at -1 PN order relative to the GR term. In practice, the dipolar contribution from the scalar field will dominate the energy loss, followed by the quadrupolar terms. Thus, we will neglect the monopole contribution and the higher PN order contributions to the dipolar radiation in our calculations. The set of parameters $\{ \pbdot, \gamma, \omegadot, r, s\}$ will be referred to as the post-Keplerian parameters from now on and will be the main set of parameters we use to constrain STTs.

\begin{table*}[t]
	\centering
	\renewcommand{\arraystretch}{0.9}
	\begin{tabular*}{6 in}{L{.8 in}| L{1.2 in} L{1.2 in} L{1.2 in} L{1.2 in}  C{1 pt}}
		&	J1738+0333	&	J1012+5307	&	J2222-0137	&	J1909-3744	&	 \\ [1 pt] 
		\hline
		\hline \\[-6pt]
		$P_b$ (d) 						&	0.3547907398724(13)		&	0.60467271355(3)		&	2.44576469(13)	&	1.533449474406(13)	&	 \\ [2 pt] 
		$x$ (s) 						&	0.343429130(17)			&	0.5818172(2)			&	10.8480239(6)	&	1.89799118	&	 \\ [2 pt] 
		$e$  							&	$3.5(1.1)\times 10^{-7}$&	$1.2(3)\times 10^{-6}$	&	0.000380967(30)	&	$1.14(10)\times 10^{-7}$	&	 \\ [2 pt]
		$m_A$ ($\msun$) 			&	1.47(7)*				&	1.64(22)*				&	1.76(6)*		&	1.47(3)*	&	 \\ [2 pt]  
		$m_B$ ($\msun$) 				&	0.181(8)				&	0.16(2)					&	1.293(25)*		&	0.208(2)	&	 \\ [2 pt] 
		$q\equiv m_A/m_B$ 				&	8.1(2)					&	10.5(5)					&	---				&	---		&	 \\ [2 pt] 
		$\pbdot^\intr$ (fs s$^{-1}$) 	&	-25.9(3.2)				&	-15(15)					&	-60(90)			&	---	&	 \\ [2 pt] 
		$\omegadot$ ($\deg$ yr$^{-1}$) 	&	---						&	---						&	0.1033(29)		&	---	&	 \\ [2 pt] 
		$s\equiv \sin\iota$  			&	---						&	---						&	0.99559			&	0.99771	&	 \\ [2 pt] 
		$r$ ($T_\odot$)							&	---						&	---						&	1.293(25) 	&	---	&	 \\ [2 pt] 
		\hline\\[-6pt]
	\end{tabular*}
	\begin{tabular*}{6 in}{L{.8 in}| L{1.2 in} L{1.2 in} L{1.2 in} L{1.2 in}  C{1 pt}}
		&	J0737-3039A	&	B1913+16	&	B1534+12	&		&	 \\ [1 pt] 
		\hline
		\hline\\[-6pt]
		$P_b$ (d) 						&	0.10225156248(5)&	0.322997448918(3)	&	0.420737298879(2)	&		&	 \\ [2 pt] 
		$x$ (s) 						&	1.415032(1)		&	2.341776(2)			&	3.7294636(6)		&		&	 \\ [2 pt] 
		$e$  							&	0.0877775(9)	&	$0.61713404(4)$		&	0.27367752(7)		&		&	 \\ [2 pt] 
		$m_A$ ($\msun$) 			&	1.3381(7)*		&	1.438(1)*			&	1.3330(2)*		&		&	 \\ [2 pt] 
		$m_B$ ($\msun$) 				&	1.2489(7)*		&	1.390(1)*			&	1.3455(2)*				&		&	 \\ [2 pt] 
		$q\equiv m_A/m_B$ 				&	1.0714(11)		&	---					&	---				&		&	 \\ [2 pt] 
		$\pbdot^\intr$ (fs s$^{-1}$) 	&	-1252(17)		&	-2398(4)			&	---			&		&	 \\ [2 pt] 
		$\omegadot$ ($\deg$ yr$^{-1}$) 	&	16.89947(68)	&	4.226585(4)			&	1.7557950(19)		&		&	 \\ [2 pt] 
		$\gamma$ (ms) 					&	0.3856(26)		&	4.307(4)			&	2.0708(5)		&		&	 \\ [2 pt] 
		$s\equiv \sin\iota$  			&	0.99974(39)		&	0.68(10)			&	0.97772			&		&	 \\ [2 pt] 
		$r$ ($\mu$s)					&	6.21(33)		&	9.6(3.5)			&	6.6(2) 	&		&	 \\ [2 pt] 
		\hline
	\end{tabular*}
	\caption[Pulsar Data]{\label{ch5:tab:pulsar_data} The Keplerian and post-Keplerian parameters measured for the systems we consider in this paper. Values with and asterisks are derived quantities assuming GR as the underlying theory. The values in parentheses represent the $1\sigma$ errors associated with each quantity. Note that the Shaprio parameter $r$ is measured in units of $T_\odot = G \msun/c^3 =  4.925490947 \mu$s for J2222-0137.
	}
\end{table*}
\subsection{Systems Considered}

Typically, PSR-WD systems provide the tightest constraints on STTs due to the predicted dipolar contribution to $\pbdot$, which has the largest effect when the binary components have very different scalar charges (as is the case with PSR-WD systems). However, one of the goals of this paper is to investigate how other post-Keplerian parameters affect the $(\alpha_0,\,\beta_0)$ parameter space. Therefore, we are also particularly interested in PSR-NS systems that have multiple post-Keplerian parameters measured to high precision. While the dipolar radiation is generally suppressed in such systems, $\omegadot$ and $\gamma$ can become very large (and therefore inconsistent with observations) due to the presence of the scalar charges $\beta_{A,B}$ and $\kappa_A$, which can take values $\sim 10^2$ and even $\sim 10^3$ in some cases~\cite{david-private}.

The set of PSRs in~\cite{Shao:2017gwu} satisfy our first criteria for PSR-WD systems with mass measurements. These PSRs include J1738+0333~\cite{Antoniadis:2012vy,Freire:2012mg}, J2222-0137~\cite{Kaplan:2014mka,Cognard:2017xyr}, J1012+5307~\cite{Lazaridis:2009kq}, and J1909-3744~\cite{Reardon:2015kba}\footnote{We exclude J0348+0432 due to the large mass ($\sim 2\msun$) of the PSR in this system. The scalar charges we are using, a subset of what is in~\cite{david-private}, only go up to 2.1$\msun$ ($2\msun$ for some EOS), and thus, the priors we use limit our exploration of the posterior in the neighbourhood of $2\msun$. This can be easily relaxed in the future if desired.}. In one way or another, there have been measurements of $\pbdot$, the companion mass $m_B$, and the mass ratio $q\equiv m_A/m_B$, thus allowing us to pin down the masses of the system. In the case of J1738+0333 and J1012+5307, there have been independent mass measurements from optical observations of the white dwarf companions, and thus, there are no correlations between any of the parameters. For J2222-0137 and J1909-3744, there exist measurements of the Shapiro parameters $r$ and $s$, which provide the extra information needed to pin down the masses. These systems also contain a diverse set of PSR masses that allow us to probe the $(\alpha_0,\,\beta_0)$ parameter space of STTs over a wide range of masses in which strong field effects like scalarization behave differently. 

In terms of PSR-NS systems, we study J0737-3039A~\cite{Kramer:2006nb,Kramer:2009zza} and B1913+16~\cite{Weisberg:2010zz}, as these PSRs have been precisely timed and there exists measurements of multiple post-Keplerian parameters. Table~\ref{ch5:tab:pulsar_data} summarizes the Keplerian and post-Keplerian parameters associated with each of these PSRs. For B1913+16 there are measurements of all 5 post-Keplerian parameters introduced in Sec.~\ref{ch5:sec:background:timing}. J0737-3039A  is part of a double PSR system, and there are measurements of all 5 post-Keplerian as well. But here there is also an independent measurement of the mass ratio from the orbital velocities of the two PSRs, providing a total of 6 parameters that can be used to place constraints.

\begin{figure*}[t]
	\centering
	\includegraphics[width=2.9in]{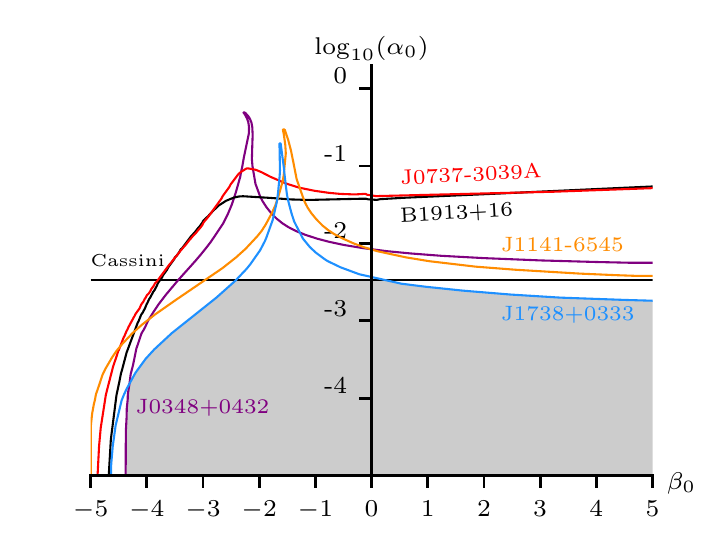}
	\includegraphics[width=2.9in]{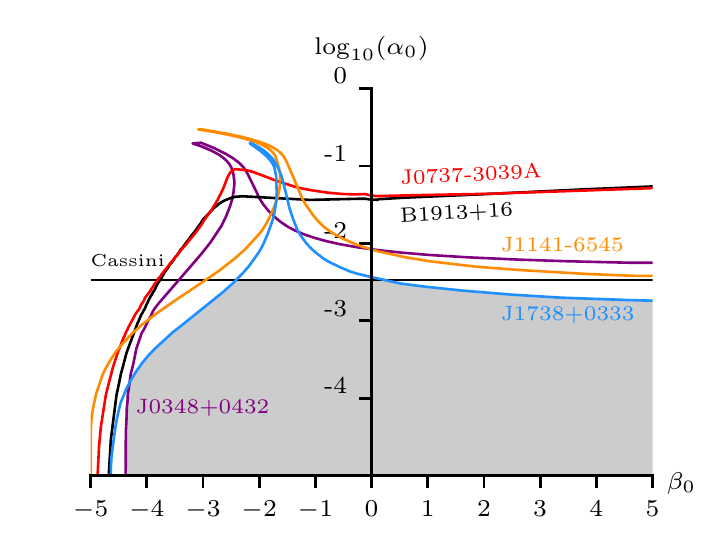}
	\caption[1$\sigma$ constarints fomr $\dot{P}_b$]{\label{ch5:fig:max_pbot} Constraints on $(\alpha_{0},\beta_{0})$ from a variety of binary PSRs on DEF theory (left) and MO theory (right) with the AP3 EOS~\cite{Akmal:1998cf}. The left panel confirms the results in~\cite{Freire:2012mg} and the right panel places the first stringent constraints with binary PSRs on MO theory. Observe that the constraints on DEF and MO theory are comparable. The horns appearing near $\beta_0 = -2$, however, are more pronounced in MO theory, because these theories are very different in this region of parameter space. The shaded gray regions are still allowed by current PSR and Solar System (Cassini~\cite{Bertotti:2003rm}) observations.
	}
\end{figure*}


\section{Constraints on STTs from binary PSRs: $1\sigma$ Constraints}\label{ch5:sec:pulsars}

In this section we will focus on placing constraints on the STT parameters $\alpha_0$ and $\beta_0$ by saturating the bounds on individual post-Keplerian parameters. As we have mentioned, some of the tightest constraints on STTs come from the lack of observed dipolar radiation in binary PSRs. In principle, if one has measured the orbital decay rate to be $\pbdot{}_{,\obs}$ up to some uncertainty $\delta \pbdot$ and found that it is consistent with the GR prediction, then Eq.~\eqref{ch5:eq:pbdot_intrinsic} can be used to place constraints on the scalar charges $\alpha_{A,B}$, and therefore, the theory parameters $(\alpha_0, \, \beta_0)$. 
The connection between $\alpha_{A,B}$ and the theory parameters, however, is dependent on the equation of state used to calculate the scalar charges. These constraints, therefore, must assume a particular EOS, and then be repeated for all possible choices. We investigate how these constraints behave when using a large number of EOSs, and how they can be improved when using measurements of other post-Keplerian parameters, considering both DEF and MO theories. 

\subsection{Constraints from $\pbdot^\intr$}\label{ch5:sec:pulsars:max}

To place constraints using measurements of $\pbdot$, we evaluate  Eq.~\eqref{ch5:eq:pbdot_intrinsic} for the entire region of parameter space investigated in~\cite{david-private}, i.e. $-5.5 \lesssim \log_{10}(\alpha_0) \lesssim 0$ and $-5 \lesssim \beta_0 \lesssim 5$. While the masses are measured to a finite precision, we evaluate the dipole term at the best fit value of the masses for the moment, and returning to this point later. Then, by determining if these predicted values of $\pbdot^{\intr}$ lie with the range $\pbdot{}_{,\obs} \pm \delta \pbdot$ we can determine if that point in parameter space is consistent or inconsistent with observations.

\begin{figure*}[t]
	\centering
	\includegraphics[width=2.9in]{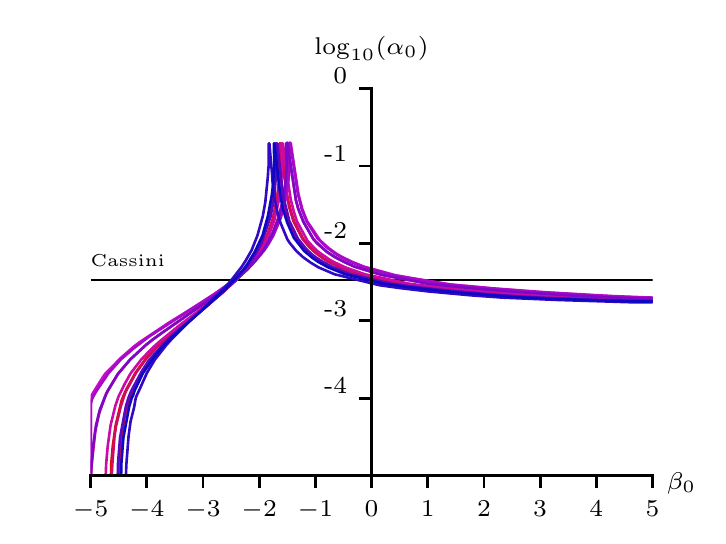}
	\includegraphics[width=2.9in]{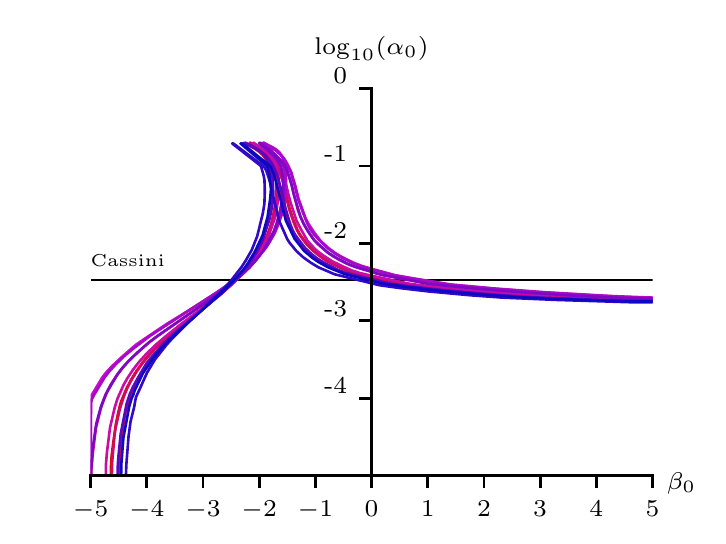}
	\caption[Effect of EOS on 1$\sigma$ constraints]{ \label{ch5:fig:max_pbot_eos}  Same as Fig.~\ref{ch5:fig:max_pbot} but using only PSR J1738+0333 and varying over 11 different EOSs~\cite{Shao:2017gwu, david-private}. While the EOSs tend to shift the curves horizontally, the relative strength of all the constraints are consistent with one another.
	}
\end{figure*}

Figure~\ref{ch5:fig:max_pbot} shows such constraints for multiple PSR systems that have accurate measurements of $\pbdot$ for both DEF theory (left panel) and MO theory (right panel) with the AP3 EOS~\cite{Akmal:1998cf}. Let us first focus on the constraints on DEF theory. These constraints confirm the results first presented in~\cite{Freire:2012mg}. The only difference between those results and the ones found here arises because the scalar charges have been here calculated much more finely in $(\alpha_{0},\beta_{0})$ space, allowing us to resolve the structure of the ``horn'' constraints more accurately. These horns arise because there are certain values of $\alpha_{0}$ and $\beta_{0}$ for which the dipole term is significantly suppressed, thus preventing any constraint. 

The constraints presented on the right panel of Fig.~\ref{ch5:fig:max_pbot} on MO theory are new. Observe that the strength of the constraints in the two theories is roughly the same. This is mostly because these theories are nearly identical to each other in the limit $\varphi \ll 1$, and therefore, they only differ substantially from each other when $\beta_0$ is very negative and/or $\alpha_0 \gtrsim 0.01$. Indeed, we see that the main difference between the constraints on the different theories is the ``horns'' that appear in Fig.~\ref{ch5:fig:max_pbot} for large values of $\alpha_0$ and $\beta_0 < 0$. 

\begin{figure*}[t]
	\centering
	\includegraphics[width=2.9in]{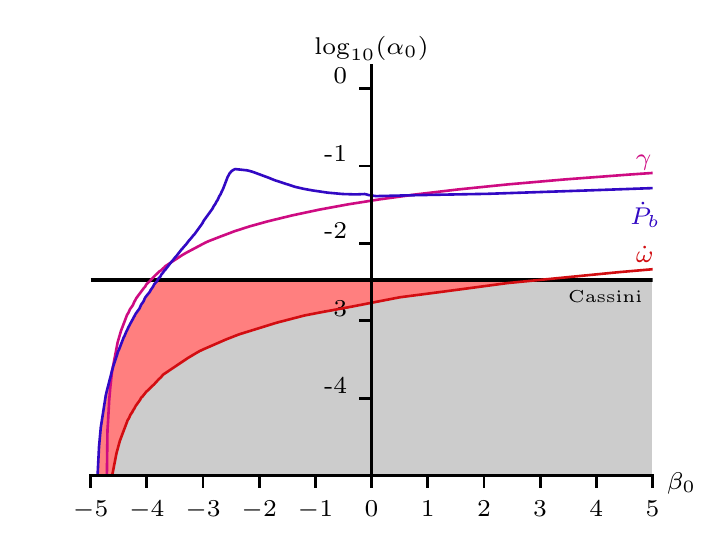}
	\includegraphics[width=2.9in]{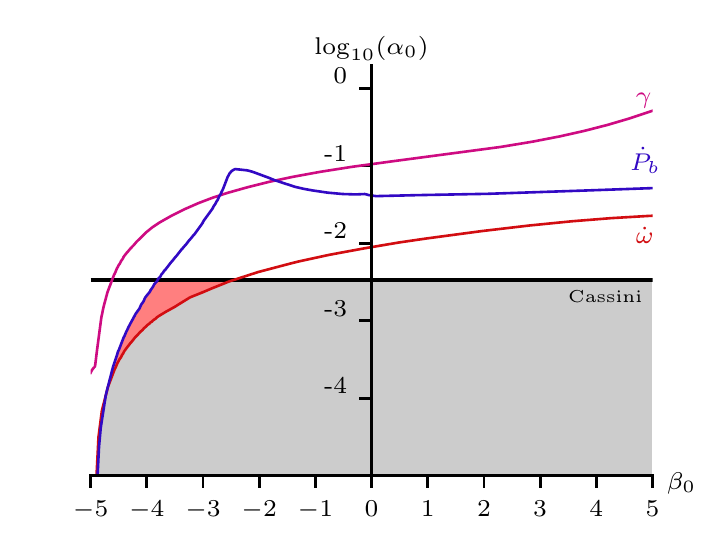}
	\caption[1$\sigma$ constraints from other post-Keplerian parameters]{ \label{ch5:fig:max_ppk_ap3}  One sigma constraints on $(\alpha_0,\,\beta_0)$ from multiple post-Keplerian parameters using B1913+16 (left) and J0737-3039 (right) and the AP3 EOS~\cite{Akmal:1998cf}. For both of these PSRs, the constraints placed from $\gamma$ and $\omegadot$ are tighter than the one placed from $\dot{P}_b$ in certain regions, with  $\omegadot$ consistently placing the tightest constraints over the entire parameter space. The shaded gray regions are those permitted by \emph{all} current constraints, while the red regions are those that have been excluded upon including other post-Keplerian parameters.
	}
\end{figure*}

The EOS affects the magnitude of the scalar charges, and thus, the mass at which spontaneous scalarization occurs, and the constraints that can be obtained from observations of $\pbdot$. Let us then repeat the analysis presented above, but this time using 11 different EOSs, namely the ones previously studied in Refs.~\cite{Shao:2017gwu, david-private}: AP3-4~\cite{Akmal:1998cf}, ENG~\cite{Engvik:1995gn}, H4~\cite{Lackey:2005tk}, MPA1~\cite{Muther:1987xaa}, MS0~\cite{Mueller:1996pm}, MS2~\cite{Mueller:1996pm}, PAL1~\cite{Prakash:1988md}, SLy4~\cite{Douchin:2001sv}, and WFF1-2~\cite{Wiringa:1988tp}.  Figure~\ref{ch5:fig:max_pbot_eos} illustrates the effect that the EOS has on the constraints placed by J1738+0333. As one can see, the EOS simply shifts the location of the horns left and right. This is because different EOS predict different masses at which scalarization kicks in, modifying the values of $(\alpha_{0},\beta_{0})$ at which the dipole term is suppressed. Aside from this modification, the variation of the EOS does not affect the overall strength of the constraints much, especially in regions where spontaneous scalarization does not occur, confirming previous results from~\cite{Freire:2012mg,Shao:2017gwu}. Although one ought to marginalize over our ignorance of the EOS, as we will discuss in the next section, the above analysis suggests that the constraints placed on DEF theory in the past~\cite{Freire:2012mg} are robust to this ignorance. 

\subsection{Constraints from other Post-Keplerian parameters}

For most systems, measurements of $\pbdot$ place the tightest constraints on STTs since the dipolar contribution to Eq.~\eqref{ch5:eq:pbdot_intrinsic} enters at -1 PN order relative to the quadrupole terms, and is therefore dominant. However, for double NS systems, the other post-Keplerian parameters can be significantly affected by the scalar charges, even though the dipolar contribution to Eq.~\eqref{ch5:eq:pbdot_intrinsic} can be negligible. Thus, using the same methods used above for $\pbdot^\intr$, we here investigate the constraints that can be placed from other post-Keplerian parameters in these double NS systems. More precisely, for any one post-Keplerian parameter $d^{i}_{\obs}$ measured with $1\sigma$ accuracy $\delta d^{i}$, we calculate the values of $(\alpha_{0},\beta_{0})$ for which the predicted value of $d^{i}_{\theory}$ lies inside  $d^{i}_{\obs} \pm \delta d^{i}$.

Figure~\ref{ch5:fig:max_ppk_ap3} shows the relative strength of the constraints on $(\alpha_0,\beta_0)$ from B1913+16 and J0737-3039 using the post-Keplerian parameters $\pbdot$, $\omegadot,$ and $\gamma$. In both cases, the observations of $\omegadot$ place significantly tighter constraints on STTs than measurements of $\pbdot$, especially for $\beta_0 < 0$. This may come as a surprise but it can be understood by the functional form of $\omegadot$. The equation for $\omegadot$ contains the higher-order scalar charge $\beta_A$, which can take on significantly large values when $\beta_0 < 0$. Typically, however, these effects are suppressed by a factor of $\alpha_0^2$ for PSR-WD systems, which is enough to make these effects completely negligible~\cite{david-private}. However, when the companion is another NS, $\alpha_B^2$ can be of order unity, making these contributions large and lead to confrontation with observations as we are seeing in Fig.~\ref{ch5:fig:max_ppk_ap3}.

These constrains, particularly the ones places by $\omegadot$, are very sensitive to the masses used in Eq.~\eqref{ch5:eq:ppk_omdot}, especially when the error is small. Consider B1913+16 in GR, for instance, in which $\omegadot$ has been measured to very high precision. The current measurements of $\omegadot$ are so precise that the total mass of the binary in GR can be determined to one part in $10^6$. However, the individual masses cannot be constrained this well from the other parameters, which means that one is not well-justified in using the best-fit masses into the equations to place the above constraints. Moreover, these plots should not be taken as a hard upper limit on $(\alpha_{0},\beta_{0})$, but rather more as a guideline of what constraints could be placed. 

There are two ways to address these types of constraints in a more consistent manner and that are not prone to issues described above. One such way is to perform an analysis similar to the one in Ref.~\cite{Esposito-Farese:2011cha} in which one calculates the theoretical values of the post-Keplerian parameters and their errors for every combination of $\alpha_0$ and $\beta_0$. Then, that point in parameter space is only excluded if there does not exist a pair of masses $(m_A, \,m_B)$ that lies in the intersection of these curves. Another method, however, involves using an MCMC to investigate these constraints and it not limited to only exploring the $1\sigma$ constraints of this section. Thus, in the next section, when we use Bayesian methods to investigate these constraints in a more informative and self-consistent manner.


\section{Bayesian inference}\label{ch5:sec:pulsars:bayesian}

Thus far we have considered the tightest possible $1\sigma$ constraints that can be placed on STTs using individual PSRs. However, we have not investigated how these constraints are affected by possible covariances between the post-Keplerian parameters, or by marginalizing over our ignorance on the EOS. Thus, in this section we will employ Bayesian methods through MCMC simulations to address these short comings. 

\subsection{Basics of Bayesian Inference}

In Bayesian statistics, given a data set $d_n$ and some hypothesis $H$ (playing the role of the theory in this case), the posterior distribution on the parameters of the hypothesis are determined by Bayes' theorem
\be
P(\vec{\lambda}|d_n, H) \eq \dfrac{ P(\vec{\lambda}| H) P(d_n|\vec{\lambda}, H)}{P(d_n|H)}\,\,,
\label{ch5:eq:bayes_theorem}
\ee
where $P(\vec{\lambda}| H)$ is the \emph{prior} probability density of the parameters $P(d_n|\vec{\lambda}, H)$ is the \emph{likelihood} of the data given the model $H$ and parameters $\vec{\lambda}$, and $P(d_n|H, I)$ is the model evidence that plays the role of a normalization factor here. Therefore, if one has the priors and the likelihood, then in principle the posterior distribution of $\vec{\lambda}$ can be calculated. In order to explore these posterior distributions efficiently, we use MCMC simulations for each data set $d_n$, corresponding to the independent observations of each PSR. 

For our purposes, $\vec{\lambda} = \{\xi_i, m_A, \,m_B, \, \log_{10}(\alpha_0), \,\beta_0\}$, and we enforce uniform priors for all parameters. The discrete parameter $\xi_i$ represents a given EOS and can take integer values from 0 to 10, with 0 corresponding to AP3 and 10 to WFF2 (ordered alphabetically). For the masses, the priors ranges are $0.01 \msun < m < 1.44 \msun$ if the mass corresponds to a WD and $1 \msun < m < 2 \msun$ if the mass corresponds to a NS\footnote{We have restricted our investigation to this range of NS masses because it becomes numerically complicated to deal with the scalar charges above this range. For our study, however, this does not limit our investigation because we do not include PSR systems with large NS masses and indeed these priors do not affect our final results.}. The STTs parameters have the prior ranges $-5.5 < \log_{10} \alpha_0 < -1$ and $-5 < \beta_0 <5$. The priors we have chosen are primarily constrained by the information we have available about the scalar charges today. Despite being able to, we \emph{do not} allow $\log_{10} \alpha_0$ to reach 0, nor do we allow the NS mass to exceed $2 \msun$. We restrict $\log_{10} \alpha_0$ simply because we find that our MCMC chains never explore these regions (as one might expect given the results of Sec.~\ref{ch5:sec:pulsars:max}. We restrict the masses of the NSs to be less than $2\msun$ because not all of the EOS we consider can achieve masses much larger than this.

For our logarithmic likelihood we use a multivariate Gaussian distribution given by
\be
\ln \mathcal{L}(\vec{\lambda}) \eq -\dfrac{1}{2} [d_{\theory}^i(\vec{\lambda}) - d_n^i] \,\mathbb{C}^{-1} _{ij}  \,[d_{\theory}^j(\vec{\lambda}) - d_n^j]\,\,,
\label{ch5:eq:logL}
\ee
where $d_n^i$ are the post-Keplerian parameters of the $n$th data set (or $n$th PSR), $d_{\theory}^i(\vec{\lambda})$ are the theoretical prediction for the same post-Keplerian parameters given the parameters $\vec{\lambda}$, and $\mathbb{C}^{-1} _{ij}$ is the inverse of the correlation matrix associated with the data set $d_n$. This likelihood has a maximum when the theoretical post-Keplerian parameters are precisely equal to their observed values and adequately handles the covariances that exist between the observed post-Keplerian parameters. The likelihood contains the theoretical predictions of the post-Keplerian parameters $d^{i}_{\theory}$, given in Eqs.~\eqref{ch5:eq:ppk_gamma}-\eqref{ch5:eq:ppk_s} and Eqs.~\eqref{ch5:eq:pbdot_intrinsic}-\eqref{ch5:eq:pbdot_quad_GR}, which in turn all depend explicitly on the scalar charges $(\alpha_{A,B},\,\beta_{A,B},\,k_A)$. For any WD companions, the charges reduce to $\alpha_B = \alpha_\infty$ and $\beta_A =  \alpha_A / \varphi_\infty \eq \beta_\infty$  since WD are weakly self gravitating. For NSs on the other hand, the scalar charges are functions of the NS mass, $\alpha_0$, and $\beta_0$ and must be solved numerically. To avoid this last step, we linearly interpolate the data from~\cite{david-private} for the scalar charges, allowing the quick computation of the likelihood over the entire prior range of the parameters. 

\begin{figure*}[t]
	\centering
	\includegraphics[width=6in]{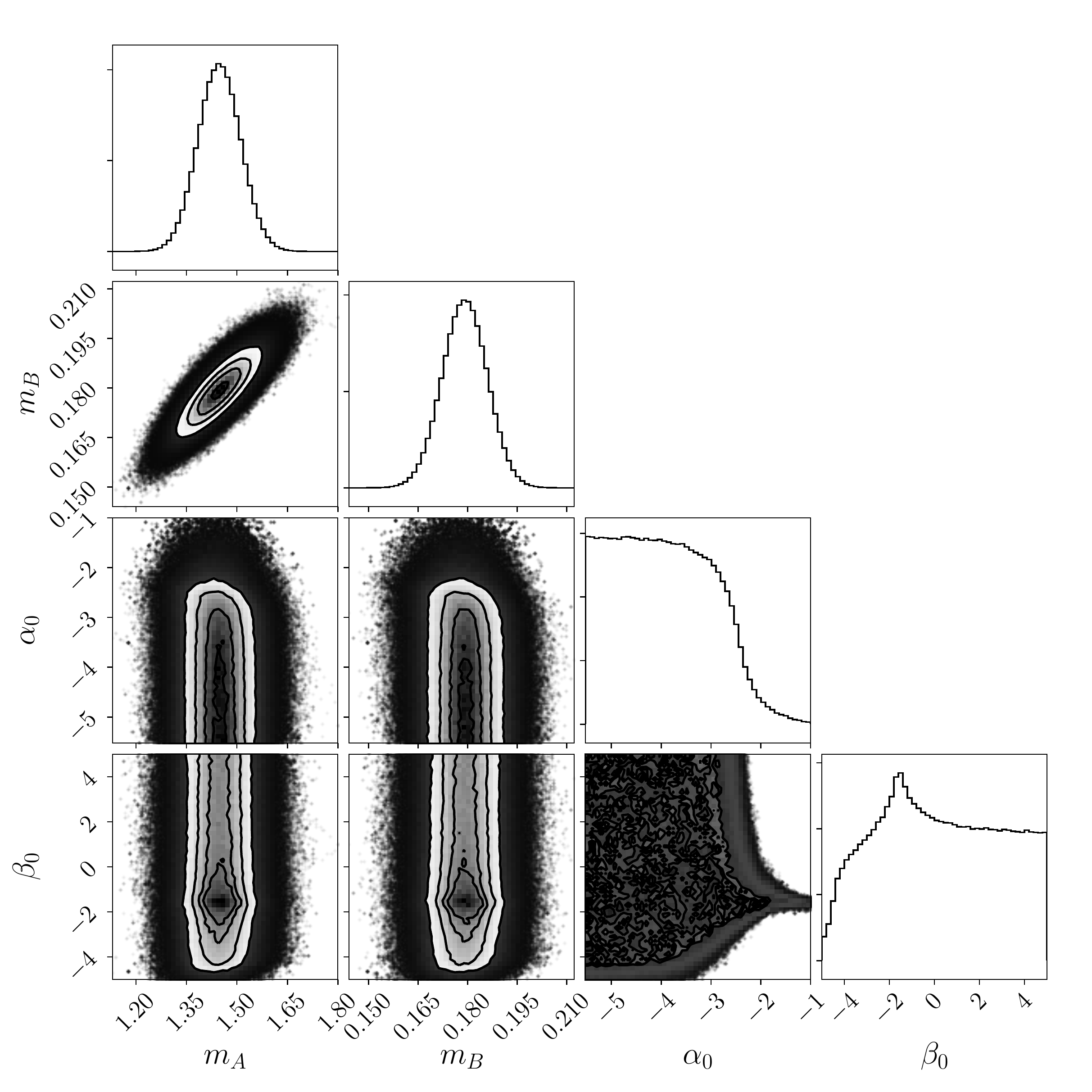}
	\caption[Marginalize Posteriors for J1738+0333]{ \label{ch5:fig:post_J1738}  Corner plot with the marginalized posteriors distributions for $(m_A,\,m_B,\,\log_{10}\alpha_0,\,\beta_0)$ from MCMC simulations using observations of post-Keplerian parameters from J1738+0333 for DEF theory.
	}
\end{figure*}

We start our MCMC chains near the binary masses predicted by GR, with EOS AP3, $\log_{10}\alpha_0 = -2$, $\beta_0 = -2$. While we start the chains near the expected peaks of the posterior distributions, the chain are allowed to explore the entire prior range, and we find that, after a burn-in phase, the chains always find these peaks regardless of where we start them. The parameter space is then explored through a proposal distribution that is equivalent to a random draw from a Gaussian distribution, with a variance that is different for each of the parameters. The variances are chosen to ensure decent acceptance ratios and autocorrelation lengths, and in practice they are different for each PSR.. 

Proposed jumps are accepted or rejected based on the Metropolis-Hastings algorithm. The Metropolis ratio for symmetric proposal distributions is just the ratio of the posterior distributions, namely 
\be
r \eq \dfrac{P(\vec{\lambda}_{new}|H)  \,\,P(d_n|\vec{\lambda}_{new}, H)}{ P(\vec{\lambda}_{old}|H)\,\, P(d_n|\vec{\lambda}_{old}, H)}\,\,,
\ee
where we used Bayes theorem from Eq.~\eqref{ch5:eq:bayes_theorem} to rewrite this ratio in terms of the priors and likelihoods. The acceptance probability is then defined via $a(\vec{\lambda}_{old},\vec{\lambda}_{new}) = \text{min}(1,r)$, which determines whether or not the proposed set of parameters is accepted by the chain or not. This is accomplished by drawing a random number $u$ between 0 and 1 and comparing it to the acceptance probability, i.e. if $a \geq u$ the jump is made, and otherwise it is not. If the proposed jump leads to a larger likelihood, the jump is always accepted, but even if the new likelihood is smaller, there is still a chance that the proposal will be accepted. We repeat this process millions of time for each PSR to ensure convergence and good exploration of the posterior distributions. 

The result of an MCMC is the joint posterior distribution on all parameters $\vec{\lambda}$. To obtain information about a single parameter, say $\beta_0$, we \emph{marginalize}, or integrate, over all other parameters $\vec{\theta}$, for instance
\be
P(\beta_0|d_n, H) \eq \int P(\vec{\lambda}|d_n, H) d\vec{\theta}\,\,.
\ee
In practice, this is equivalent to creating a histogram of the likelihood evaluations over the parameter that is being marginalized.

\subsection{Bayesian Results}

Figure~\ref{ch5:fig:post_J1738} shows the marginalized posteriors on the parameters $(m_A,\,m_B,\,\log_{10}\alpha_0,\,\beta_0)$ for J1738+0333. The marginalized posteriors on $\xi_i$ for the EOS is uniform, meaning that the MCMC showed no preference toward any particular EOS and thus we do not show it because it is uninformative. We recover mass distributions that are consistent with observations and the predictions of GR. The posteriors on $\alpha_0$ and $\beta_0$ behave as one would expect. Lower values of $\alpha_0$ (towards GR) are preferred and the posteriors on $\beta_0$ have a nearly identical form as the constraints placed by $\pbdot$ in the previous section. The latter makes sense since we are not enforcing Solar System priors on the parameters, and thus, the MCMC spends more time exploring regions that are less constrained, i.e. near $\beta_0 = -2$. 

\begin{figure*}[t]
	\centering
	\includegraphics[width=2.9in]{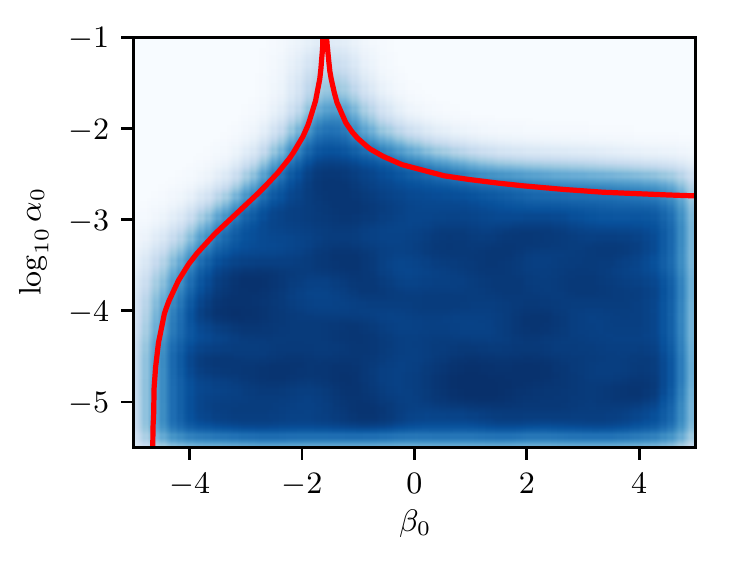}
	\includegraphics[width=2.9in]{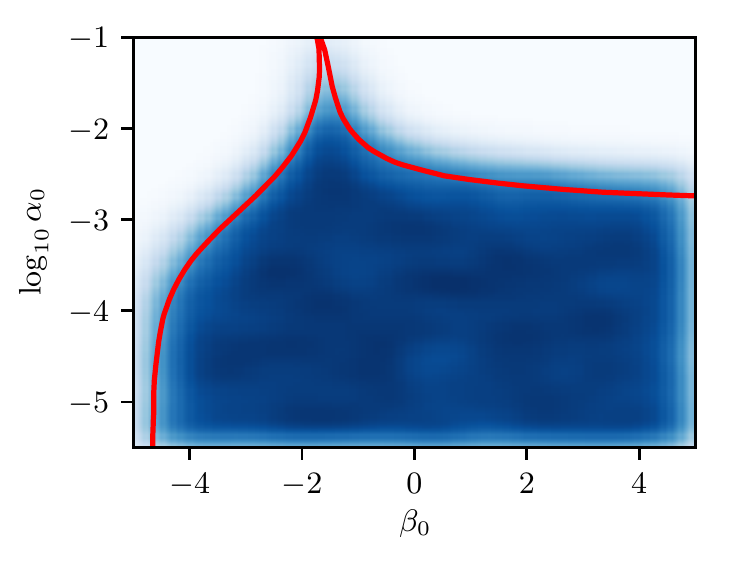}
	\caption[Marginalize posteriors on $(\alpha_0,\,\beta_0)$]{ \label{ch5:fig:post_joint}  Marginalized posteriors distributions for $(\log_{10}\alpha_0,\,\beta_0)$ from MCMC simulations of the post-Keplerian parameters extracted from J1738+0333, using DEF theory (left) and MO theory (right), with the AP3 EOS. We overlay the constraints found in Sec.~\ref{ch5:sec:pulsars:max} in red for comparison. One can see that the MCMCs predict marginalized posteriors that are consistent with the $1\sigma$ constraints.
	}
\end{figure*}

The joint posterior on $\alpha_0$ and $\beta_0$ are presented in Fig.~\ref{ch5:fig:post_joint}, where we also include an overlay of the constraints placed from the previous section. This figure is really a heat plot of the posterior distribution, with brighter (darker) colors representing low (high) posterior probabilities. Therefore, the dark regions are unconstrained by our Bayesian analysis, while the bright regions are excluded. We see great consistency between the Bayesian constraints and the constraints from the previous subsection.

Thus far we have only discussed results involving single PSRs with associated data $d_n$, so let us now combine the constraints from multiple observations in two different ways. The first involves constructing a new log likelihood that contains information from \emph{all} of the PSRs we consider. This is equivalent to simply adding the log likelihoods (multiplying the actual likelihoods) associated with each individual PSR and running a new MCMC with a new parameter vector given by $\vec{\lambda}_{\text{total}} = \{\sum_{n} m_{A,n }, \sum_{n} \,m_{B,n}, \, \xi_{i}\,, \alpha_0,\, \beta_0\}$ where $n$ stands for the $n$th binary system. One would then marginalize the likelihood over all parameters that are not $(\alpha_{0},\beta_{0})$ to find the new joint posterior. This is quite possible but it requires us to effectively recompute the posteriors we have already found. 

Another way to combine information is to carefully ``stack'' the joint posteriors on the parameters $\vec{\sigma} \eq (\xi_i, \,\alpha_0,\,\beta_0)$ that we have already calculated\footnote{It is impossible to stack posteriors on single parameters simply because the operations of integration and multiplication do not commute. Thus, we are only able to stack marginalized posteriors if we have marginalized over parameters that are not shared in different observations, like the masses in this case.}. In this case the total posterior distribution is just the product of the individual posteriors. To do this we start with Bayes theorem in Eq.~\eqref{ch5:eq:bayes_theorem} with total data $D$ from $N$ systems, and we note that the priors are identical with the likelihood following  the relation
\be
\dfrac{P(D|\vec{\sigma})}{P(D|H)} \eq \prod\limits_{n=1}^N \dfrac{P(d_n|\vec{\sigma}, H)}{P(d_n|H)} \eq \prod\limits_{n=1}^N \dfrac{P(\vec{\sigma}|d_n, H)}{P(\vec{\sigma}|H)}\,\,,
\label{ch5:eq:joint_like}
\ee
Putting this information back into Bayes theorem for the total posterior we find the relation
\be
P(\vec{\sigma}|D, H) \eq P(d_n|\vec{\sigma}, H)^{1-N} \prod\limits_{n=1}^N P(\vec{\sigma}|d_n, H)\,\,.
\label{ch5:eq:stacked_post}
\ee
Therefore, we are able to take the individual joint posteriors on $\vec{\sigma}$ that we have already calculated and multiply them in this manner to find the total posterior we would have found had we done a new MCMC simulation with all PSRs included.

\begin{figure*}[th!]
	\centering
	\includegraphics[width=2.9in]{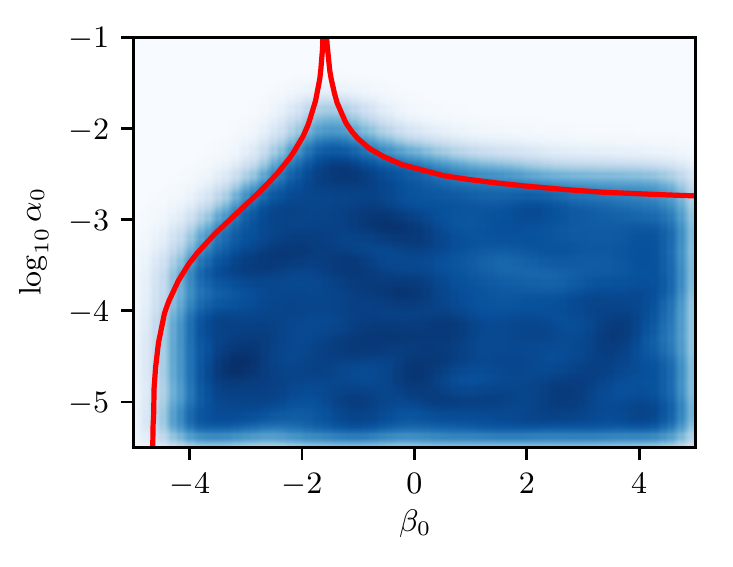}
	\includegraphics[width=2.9in]{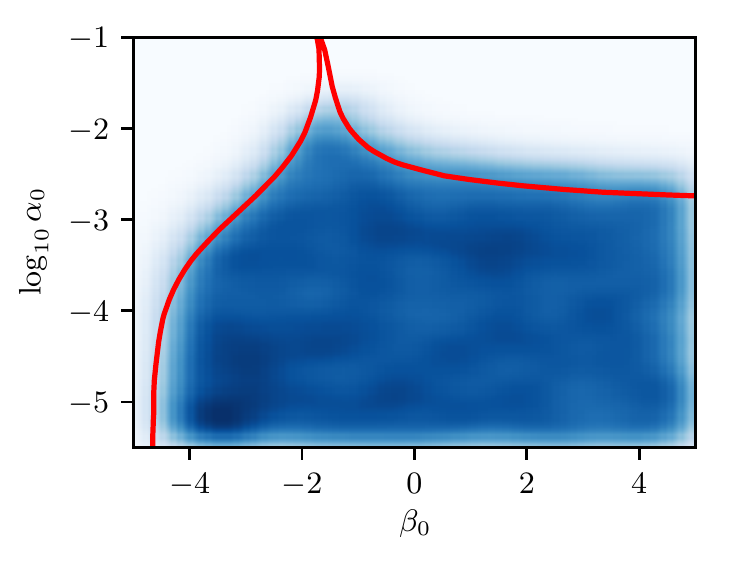}
	\caption[Stacked posteriors on $(\alpha_0,\,\beta_0)$]{ \label{ch5:fig:post_stacked}  Same as Fig.~\ref{ch5:fig:post_joint} but stacked over information from the PSRs in Table~\ref{ch5:tab:pulsar_data} for DEF theory (left) and MO (right) theory using the AP3 EOS. Again, we overlay the constraints found in Sec.~\ref{ch5:sec:pulsars:max} for J1737+0333 in red for comparison. Since J1737+0333 places the tightest constraints in Sec.~\ref{ch5:sec:pulsars:max} it makes sense that the stacked posterior is consistent with these constraints. 
	}
\end{figure*}

Figure~\ref{ch5:fig:post_stacked} shows the total joint posterior on $(\alpha_0,\,\beta_0)$, marginalized over the EOS, after combining the posteriors from the PSRs in Table~\ref{ch5:tab:pulsar_data} via Eq.~\eqref{ch5:eq:stacked_post}. The overall effect of this ``stacking'' of the posteriors does not produce a significant effect but it does visibly reduce the probability of the larger values of $\alpha_0$ near $\beta_0 \sim -2$. This results makes sense when compared to results found in Sec.~\ref{ch5:sec:pulsars:max}. Each PSR tends to place different constraints in this region of parameter space, and the Bayesian methods we have employed here allow us to take these multiple constraints into consideration simultaneously.

Rather than assuming flat priors for the STTs parameters we can enforce constraints that have been placed from other tests, like Solar System tests, through different priors. The Solar System constraint on the post-Keplerian parameters appear in Eqs.~\eqref{ch5:eq:ppn-gamma}-\eqref{ch5:eq:alpha_inf2}, and numerically, these constraints turn into $\alpha_\infty \leq 0.003391$. As we have pointed out in Sec.~\ref{ch5:sec:background}, this constraint on $\alpha_\infty$ technically translates to different constraints on $\alpha_0$ in each theory. However, for practical purposes the constraints on $\alpha_0$ are identical in both DEF and MO theory because of the small magnitude of $\alpha_\infty$. 

\begin{figure*}[t]
	\centering
	\includegraphics[width=2.9in]{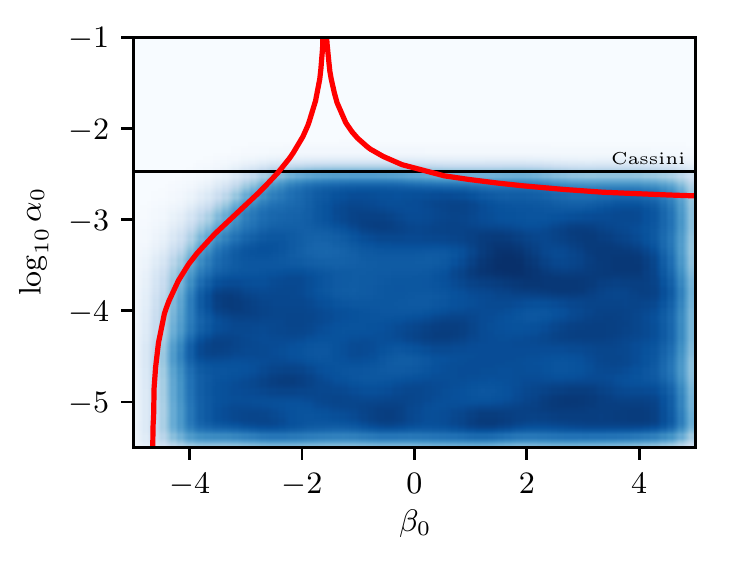}
	\includegraphics[width=2.9in]{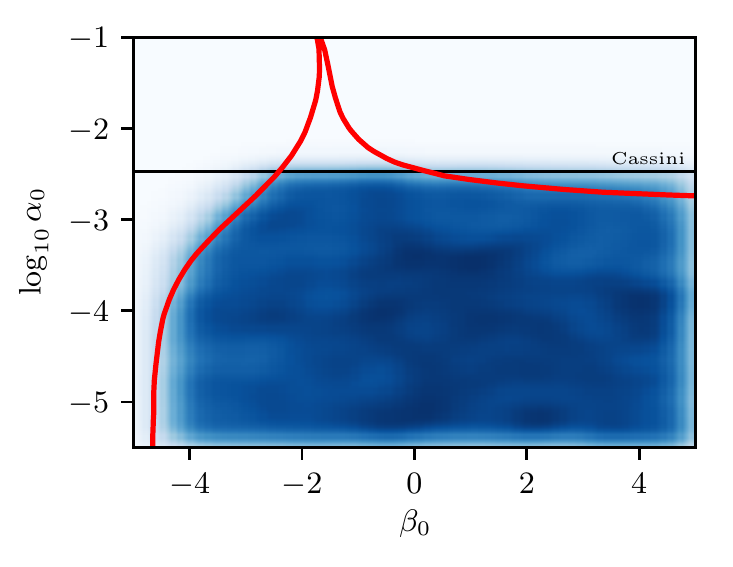}
	\caption[Stacked posteriors on $(\alpha_0,\,\beta_0)$ with Solar System priors]{ \label{ch5:fig:post_stacked_ss}  Same as Fig.~\ref{ch5:fig:post_stacked} but where we have enforced Solar System constraints on $\alpha_0$ through priors.
	}
\end{figure*}

The effect of more stringent priors is to tighten the posteriors on all parameters. These priors effectively remove the posterior distributions on $\alpha_0$ larger than $0.003391$ that we have presented thus far and therefore reduce the parameter space that the MCMC explores. This results in tighter posteriors for the masses of the systems, as well as more stringent constraints on $(\alpha_{0},\beta_{0})$ as shown in Fig.~\ref{ch5:fig:post_stacked_ss}.


\section{Conclusion}\label{ch5:sec:conclusion}


We have investigated the constraints that binary PSRs can place on the parameters $\alpha_0$ and $\beta_0$ that appear in scalar-tensor theories, particular in DEF theory and in MO theory. We show for the first time in MO theory the type of constraints that one can place on STTs using observed orbital period decay $\pbdot$. Then, using other post-Keplerian parameter we demonstrate that for certain systems, like double NS systems, even tighter constraints can be place on the $(\alpha_0,\,\beta_0)$ parameter space of STTs. We find consistency with these results when we use Bayesian methods to investigate the posterior distributions of the parameters and find that the dominant effects the dictate the overall constraints is indeed $\pbdot$.

This work has been one of the first attempts to apply an MCMC to explore the constraints on STT parameter space from binary PSRs, and therefore, it should be considered a first step towards a more in depth Bayesian study. For example, we have only used the quoted values of the post-Keplerian parameters, along with the correlations, to construct our likelihood functions and explore the posterior distributions of $\alpha_0$ and $\beta_0$. Such a method introduces systematic errors because we are not able to incorporate covariances between the normal Keplerian parameters, even though these parameters are measured to extremely high precision. Our analysis also potentially suffers from having a limited number of data points, i.e. the post-Keplerian parameters here.

A more complete study, and one we hope to perform in the future, would involve using more sophisticated MCMC techniques like parallel tempering, to allow us to explore potentially multi-modal posteriors, or like those implemented in the emcee\_hammer package available for Python. Moreover, one could also re-analyze the PSR data to incorporate the TOAs directly into the likelihood and allow an MCMC to perform the fits for the Keplarian and the post-Keplerian parameters simultaneously, essentially performing the computations that timing packages like TEMPO and TEMPO2 handle. This would allow one to easily fit the TOAs given any gravitational theory, like GR and STTs, that have a well developed timing formula. The benefit of handling the TOA directly is that it removes any systematics and inconsistencies that arise in our study, i.e. trying to analyze data that has already been processed. 

Another possible avenue for future work is to consider more complicated binary pulsar systems in a Bayesian approach to test GR. A particular interesting system is the triple system recently observed~\cite{}. This observation allows for a measurement of a parameter that constraints a certain type of SEP violation. This parameter depends on the scalar charges in STTs, as well as on the component masses of inner binary. One could thus repeat the analysis presented here for this system, marginalizing over the component masses and over the EOSs to obtain new stringent constraints on DEF and MO theory.   


\acknowledgements\label{ackno}
We would like to thank Norbert Wex for useful discussions on constraining STTs and Neil Cornish and Travis Robson for useful discussions on Bayesian statistics. DA would like to thank Paulo Freire and the Max-Planck-Institut f\"ur Radioastronomie for their hospitality during part of this work. NY and DA acknowledge support from NSF grant PHY-1759615 and NASA grants NNX16AB98G and 80NSSC17M0041.


\bibliography{thesis}
\end{document}